# Graphite Nanoeraser

Ze Liu,[1,2] Peter. Bøggild,[3*] Jia-rui Yang,[1,2] Yao Cheng,[4,2] Francois Grey,[2] Yi-lun Liu,[1,2] Li Wang,[5] Quan-shui Zheng,[1,2,5*]

**We present here a method for cleaning intermediate-size (5~50nm) contamination from highly oriented pyrolytic graphite. Electron beam deposition causes a continuous increase of carbonaceous material on graphene and graphite surfaces, which is difficult to remove by conventional techniques. Direct mechanical wiping using a graphite nanoeraser is observed to drastically reduce the amount of contamination. After the mechanical removal of contamination, the graphite surfaces were able to self-retract after shearing, indicating that van der Waals contact bonding is restored. Since contact bonding provides an indication of a level of cleanliness normally only attainable in a high-quality clean-room, we discuss potential applications in preparation of ultraclean surfaces.**

Surface contamination of graphite and graphene is a common problem in micro- and nanotechnology. Graphene can be isolated by mechanical exfoliation[1] as well as chemical methods, surface segregation and decomposition strates[2,3], due to the extreme sensitivity of the physical, chemical and electrical properties to the environmental conditions, the ability to clean the graphene is an essential prerequisite for most applications[4,5,6,7].

Smaller absorbents such as individual molecules and atoms on graphite/graphene surface can be cleaned by a chemical treatment[7] or by high electrical current[8] or high temperature in Ar/H2 environment[4,5], while much larger microsize structures can be removed by a strong airflow, by ultrasonic vibration[9] or by mechanical peeling (see reviews in Ref. 10). There are few demonstrations of cleaning of clusters, agglomerations, particles and other surface formations in the intermediate (5-50nm) range. Moreover, the heat-based methods cannot easily remove high-temperature melting point materials. It is well known from tribology that mechanical scraping of a surface using for instance a nanowire or a sharp tip[11,12], can reduce the amount of intermediate-size contamination, by simply displacing the surface structures. With the recent surge in graphite/graphene research and emerging applications, methods for reducing the amount of unwanted particles and structures on graphene surfaces are needed. We here explore the concept of mechanical cleaning of highly oriented pyrolytic graphite (HOPG), which is well known to accumulate amorphous carbon material during inspection inside a scanning electron microscope (SEM), due to the decomposition of residual organic molecules present in the chamber even at vacuum, as well as molecules residing on the surfaces[13]. Such films of contamination are considered to be highly difficult to remove[14,15]; for instance, freestanding nanowires formed by electron beam deposition do not break at the base upon mechanical stress, indicating a very strong connection to the underlying substrate. In this work, electron beam deposited material represents a worst-case example of the ubiquitous and hard-to-remove organic contamination encountered in visual as well as electrical characterization of nanoscale materials including graphite and graphene.

[1]Department of Engineering Mechanics, Tsinghua University, Beijing 100084, China. [2]Center for Nano and Micro Mechanics, Tsinghua University, Beijing 100084, China. [3]Mikroelektronik Centret, Technical University of Denmark, DK-2800 Lyngby, Denmark. [4]Department of Engineering Physics, Tsinghua University, Beijing 100084, China. [5]Institute of Advanced Study, Nanchang University, Nanchang 330031, China. Correspondence and requests for materials should be addressed to Q.S.Z. (email: zhengqs@tsinghua.edu.cn ), or to P.B (email: peter.boggild@nanotech.dtu.dk )

Earlier work demonstrated the self-retracting behavior of graphite flakes sheared from $SiO_2$ film-covered graphite islands.[16]. We show here that the graphite surface over time becomes visibly (in the SEM) contaminated by amorphous carbon, and that the surface can be cleaned well enough to recover the self-retracting behavior. Since the self-retraction behavior is based on van der Waals forces[16,17], and thus is dependent on intimate contact between the two graphite planes, we suggest that the cleaning method might approach atomic-level cleanliness., thus restoring the intimate contact of multiple pristine graphene flakes.

## Results

*Wiping Process.* In our experiments, the HOPG samples purchased from Veeco (ZYH Grade) are used. We first mechanically exfoliated HOPG with adhesive tape[18] to obtain a clean surface, and then fabricated graphite islands with different size as previously reported in Ref. 16 (See Methods for more details on sample preparation). Fig. 1 shows the as-fabricated graphite islands, with ~110 silicon dioxide on top of each island.

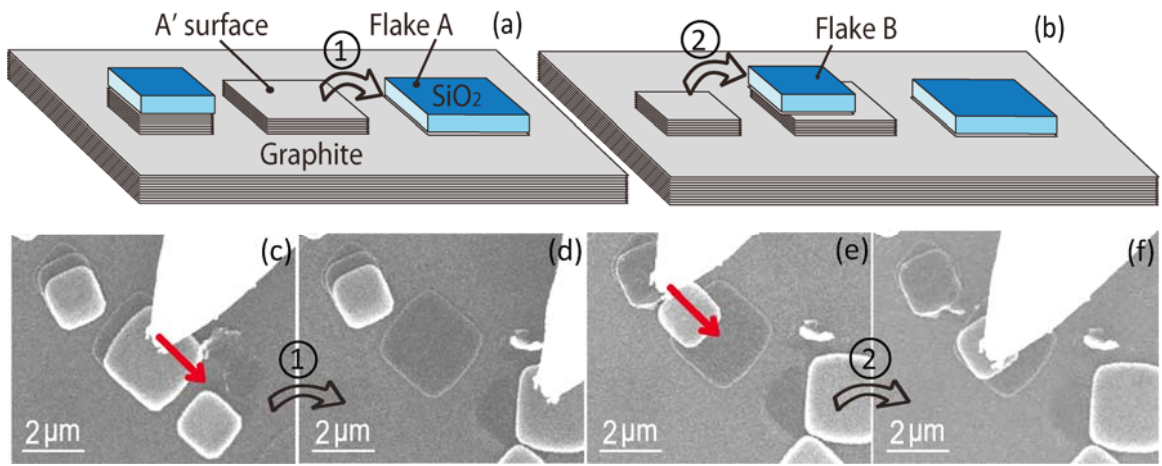

**Figure 1 | Experiments on micro-flakes** Panels (a-b) illustrate the experimental process, where a graphite/SiO$_2$ flake (A) was first slided off the graphite island and generated a fresh graphite surface (A'), then a graphite/SiO$_2$ flake (B) was slided off a smaller graphite island and shifted onto the surface A'. A neccessary condition to make the above process possible is that the spacing between the two islands is shorter than the size of Flake B, although this process was not always realizable. The SEM images (c-f) show the real micro-manipulation process with (c-d) corresponding to (a), and (e-f) to (b).

As shown in the illustrative Fig. 1a-b, we first slide a graphite/SiO$_2$ flakes (Flake A) off a micro SiO$_2$-covered graphite island by using a micromanipulator, so a fresh graphite surface below (A' surface) is exposed. We then slide a graphite/ SiO$_2$ flake (Flake B) away from a neighboring smaller SiO$_2$-covered graphite island until it attached onto the exposed (A' surface) platform. The real micro-manipulation observed in-situ in a SEM can be seen from a movie (Supplementary Movie 1) or the selected frames Fig. 1c-f of this movie, which have been rotated 180 degrees to keep according with Fig. 1a-b. Dorp et al[13] reviewed the electron beam induced deposition and concluded that the deposition rate may depend on beam current, acceleration voltage as well as concentration and distribution of residual organic material inside the chamber. In our experiments, we scanned the platform surface (A') every ~60s and observed a clear increase in surface contamination. After about 5 minutes exposure (Fig. 2a), we started to wipe the platform using Flake B. The details can be seen from the Supplementary Movie 2 or the selected frames Fig. 2b-c. By comparing the insets in Fig. 2a and Fig. 2d, which are normalized gray values along the colored lines, it can be clearly seen that the wiping process is indeed efficiency.

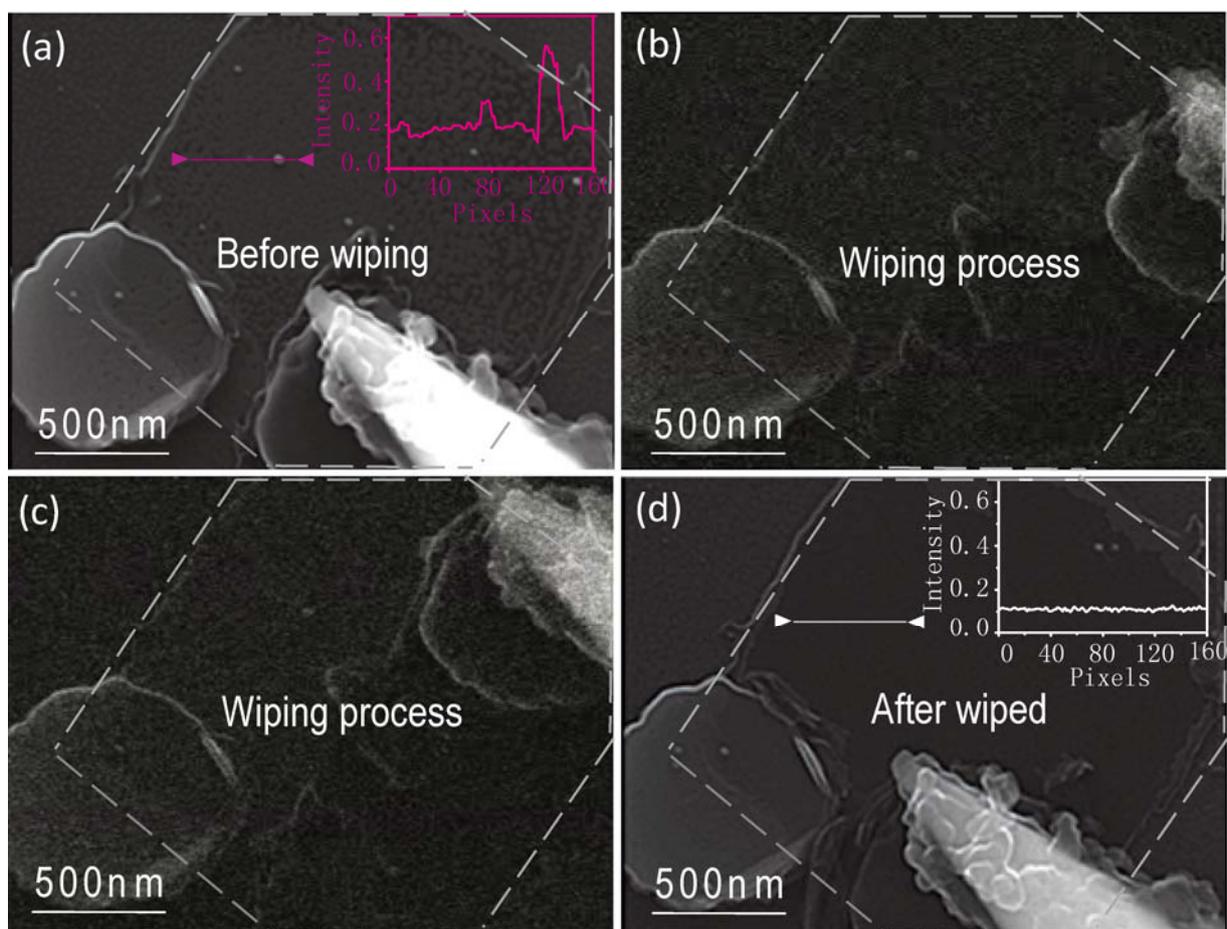

**Figure 2 | Erasing processes with gaphite/SiO2 micro-flakes in SEM** (a) The larger graphite platform (A') outlined by the dashed-line after exposed to electron beam about 5 minutes, the inset shows normalized gray values along the colored line, indicating the levels of carbon deposition. (b-c) The selected frames from Movie 3 show that a smaller graphite/$SiO_2$ was wiping on the larger graphite platform (A'). (d) After erasing, the graphite surface becomes clean again, as clearly indicated by the inset.

***Self-retractability.*** Recently we reported our discovery of the self-retraction phenomenon that after sheared a graphite/$SiO_2$ flake partly out of a $SiO_2$-cover graphite island and then released it; the flake could quickly retract back onto the platform created from the shear[16]. This phenomenon is expected to be highly dependent on intimate contact between the two graphite surfaces[9], and any contamination would strongly decrease the Van der Waals forces responsible for the self-retraction behavior. Prior to cleaning, self-retraction of the partly contacted smaller flake B onto the larger platform A', immediately after the former was attached onto the latter, was never observed. However, after the flake B had been used to clear away carbonaceous contamination from A', we pushed B partly out of A' using a micro-probe and then released. In the experiment we found the flake B was retractable! This observation clearly indicates that the surface is cleaned very efficiently, and the Van der Waals contact bonding between the smaller flake and the bigger platform is again achieved, as judged by the strong interaction between the surfaces. The detailed process can be seen from supplementary Movie 3.

In addition, we also used the electron beam-induced carbon deposition to "glue" the smaller flake B to the apex of a microprobe, analogous to soldering of microelectrodes[19]. We tried to pull off the flake B through the glued microprobe from the platform and found it quite difficult. We then further slid the flake B to an edge of the platform until it was partly suspended from the platform. We found it easy to lift B off the platform (See Supplementary Movie 4) within almost a short carbon deposition time, which is agree well with the fact that the Van der Waals force between graphite layers is proportional to their contact area. Finally, we shifted this flake B onto another larger platform to clean the surface. We suggest that such a 'nanoeraser' probe could be used to clean critically important surface areas on graphene and graphite, and possibly also structures like hexagonal Boron Nitride and mica in an easy, direct manner.

## Discussion

We present an easy method to clean the surface of highly oriented pyrolytic graphite (HOPG). Based on the self-retraction motion of graphite flakes reported by Zheng et al[16], the 'nanoeraser' concept is verified within a scanning electron microscope. We also exploit the electron beam-induced carbon deposition in SEM to fabricate a 'nanoeraser' probe. When mounted on a micromanipulator the nanoeraser is a unique, local cleaning device, which allows to selectively clean graphene and graphite surfaces with nanometer accuracy. We anticipate could be useful for investigating the effects of surface contamination, which are critically important for graphene applications. Micromanipulators are now standard equipment for SEM, hence the 'nanoeraser' could provide a simple, convenient solution to the problem of not only SEM-induced contamination, but also a range of other types of hard-to-remove surface contamination which may occur in the graphite/graphene fabrication or transfer process.

Graphene is an exceptional material for the chemical sensor to resolve individual gas molecules[20]. Chemical doping accumulated on the operation surface leads to the degrading problem as usual encountered for all kinds of the semiconductor gas sensors[21]. Removing the surface adsorbates by a wiper-like action, the reported nanoeraser provides a total solution, reversibly to activate the sensing area.

The 'nanoeraser' may also be used as a portable parallel plate electrode to drive Micro/Nano devices in lab study, by replacing the silicon dioxide with a conducting layer. The 'nanoeraser' may also supply as a 'graphene pen' to deposit graphene sheets on surfaces through friction[22], which may be tuned to overcome the weak Van der Waals interaction between graphene layers. Such behavior was especially observed when erasing rough surfaces. We also anticipate that the nanoeraser could be used to clean other smooth material surfaces such as silicon (in Ultra-high Vacuum) and mica, in situations where other cleaning methods such as chemical methods or high temperature evaporation are impractical or potentially damaging.

## Methods

**Sample preparation.** The method to prepare the islands in the experiments is similar to that reported in Ref. 16. After the freshly HOPG surface is obtained with adhesive tape[18], a thin silicon dioxide ($SiO_2$) film is deposited on the surface by plasma enhanced chemical vapor deposition (PECVD). The islands are defined by electron beam lithography using the positive resist (ZEP520) as an etch mask for the subsequent reactive ion etching with mixed gases trifluoromethane ($CHF_3$) and argon (Ar). After the $SiO_2$ film is etched away, the graphite layers beneath it is further etched about 2 minutes by substituting the reaction gas with oxygen, leading to structures as shown in Fig. 1 shows.

***In situ* experiments in SEM.** A FEI Quanta 200F scanning electron microscope operating in high vacuum conditions was used with a 30 keV acceleration voltage and a spot size of 3 nm. A Kleindieck mm3A micromanipulator with a 5 nm lateral accuracy was used for the manipulation. The experiments on each island were carried by moving a tungsten microprobe with suitable stiffness and tip apex prepared by AC Electrochemical Etching. The operating process was monitored and recorded in either images or in situ videos.

## Acknowledgments

Q.S.Z. acknowledges the financial support from NSFC through Grant No. 10832005, the National Basic Research Program of China (Grant No. 2007CB936803), and the National 863 Project (Grant No. 2008AA03Z302).

## Author contributions

Z.L. prepared the samples and carried out the experiments. Q.S.Z. and P.B. designed the project and guided the research. P.B., Z.L., and Q.S.Z. wrote the paper. J.R.Y., Y.C., F.G., Y.L.L. and W.L. involved many discussions.